\documentclass[%
reprint,
amsmath,amssymb,
aps,
prl,
superscriptaddress
]{revtex4-1}

\usepackage{csquotes}
\usepackage{graphicx} 
\usepackage{hyperref}
\usepackage[caption=false,]{subfig}
\usepackage{color}
\usepackage{xr}
\usepackage[normalem]{ulem}

\newcommand{\mum}{\mu \textnormal{m}}
\newcommand{\muM}{\mu \textnormal{M}}
\newcommand{\nM}{\textnormal{nM}}
\newcommand{\nm}{\textnormal{nm}}
\newcommand{\s}{\textnormal{s}}
\newcommand{\omon}{\omega_\textnormal A}
\newcommand{\omoff}{\omega_\textnormal D}

\usepackage{tikz}

\begin{document}  

\title{Limited Resources Induce Bistability in Microtubule Length Regulation}

\author{Matthias Rank}
\thanks{These authors contributed equally}
\affiliation{Arnold-Sommerfeld-Center for Theoretical Physics and Center for NanoScience, Ludwig-Maximilians-Universit\"at M\"unchen, Theresienstra\ss Ÿe 37, 80333 M\"unchen, Germany}

\author{Aniruddha Mitra}
\thanks{These authors contributed equally}
\affiliation{B CUBE -- Center for Molecular Bioengineering and Center for Advancing Electronics Dresden (cfaed), Technische Universit\"at Dresden, Arnoldstra\ss e 18, 01307 Dresden, Germany}
\affiliation{Max Planck Institute of Molecular Cell Biology and Genetics, Pfotenhauerstra\ss e 108, 01307 Dresden, Germany}

\author{Louis Reese}
\affiliation{Department of Bionanoscience, Kavli Institute of Nanoscience, Faculty of Applied Sciences, Delft University of Technology, Van der Maasweg 9, 2629 HZ Delft, The Netherlands
}

\author{Stefan Diez}
\email{stefan.diez@tu-dresden.de}
\affiliation{B CUBE -- Center for Molecular Bioengineering and Center for Advancing Electronics Dresden (cfaed), Technische Universit\"at Dresden, Arnoldstra\ss e 18, 01307 Dresden, Germany}
\affiliation{Max Planck Institute of Molecular Cell Biology and Genetics, Pfotenhauerstra\ss e 108, 01307 Dresden, Germany}
	
\author{Erwin Frey}
\email{frey@lmu.de}
\affiliation{Arnold-Sommerfeld-Center for Theoretical Physics and Center for NanoScience, Ludwig-Maximilians-Universit\"at M\"unchen, Theresienstra\ss Ÿe 37, 80333 M\"unchen, Germany}

\begin{abstract} 
	{ 
		The availability of protein is an important factor for the determination of the size of the mitotic spindle. Involved in spindle-size regulation is kinesin-8, a molecular motor and microtubule (MT) depolymerase, which is known to tightly control MT length. 
		Here, we propose and analyze a theoretical model in which kinesin-induced MT depolymerization competes with spontaneous polymerization while supplies of both tubulin and kinesin are limited. 
		In contrast to previous studies where resources were unconstrained, we find that, for a wide range of concentrations, MT length regulation is bistable. We test our predictions by conducting \emph{in vitro} experiments, and find that the bistable behavior manifests in a bimodal MT length distribution. 	
	}
	{}
	{}
\end{abstract}

\maketitle

The absolute and relative abundance of particular sets of proteins is important for a wide range of processes in cells. For example, during \emph{Xenopus laevis} embryogenesis, importin $\alpha$ becomes progressively localized to the cell membrane~\cite{Wilbur2013}. As a consequence of importin's depletion from the cytoplasm, the protein kif2a escapes inactivation, and decreases the size of the mitotic spindle. Similarly, formation of the mitotic spindle reduces the concentration of free tubulin dimers, the building blocks of microtubules (MTs). Thus, up to $60\%$ of all tubulin heterodimers~\cite{Borisy1967a,Borisy1967b} may be incorporated into the spindle~\cite{Good2013}. In addition, it has been shown \emph{in vivo} and \emph{in vitro} that both spindle size~\cite{Good2013,Hazel2013} and the lengths of its constituent MTs~\cite{Winey1995} scale with cytoplasmic volume. 

Assembly and disassembly of MTs are regulated by a set of proteins that interact with the plus ends of protofilaments~\cite{Howard2003,Howard2007}. One of these factors, the molecular motor kinesin-8, acts as a depolymerase~\cite{Varga2006,Howard2007}. As a consequence, spindle size increases in its absence~\cite{Goshima2005}, and decreases upon overexpression of the protein~\cite{Stumpff2008}. Moreover, the kinesin-8 homolog Kip3 from \emph{Saccharomyces ce\-re\-visiae} has been shown to depolymerize MTs in a length-dependent fashion~\cite{Varga2006,Varga2009}. This is facilitated by a density gradient on the MT, caused by the interplay between the processive motion of Kip3 along the MT and its depolymerase activity at the plus end, which effectively enables the MT to \enquote{sense} its own length~\cite{Varga2009,Reese2011}. In combination with spontaneous MT polymerization, the Kip3 gradient leads to a length regulation mechanism~\cite{Melbinger2012,Reese2014}. 

Here we explore the combined effect of limited resources and Kip3-induced depolymerization on the length regulation of MTs. As seen in theoretical studies on the collective motion of molecular motors, resource limitation affects the density profile on the MT: Regions of low and high motor density separate, as a localized domain wall emerges on the MT~\cite{Lipowsky2001,Cook2009,Greulich2012,Ciandrini2014}. 
This is a direct result of resource limitation, and does not rely on the existence of a motor density gradient, as necessary for domain wall localization in the presence of unlimited resources~\cite{Parmeggiani2003,Parmeggiani2004,Leduc2012,Subramanian2013}.
So far, most work on the role of limited resources has focused on single components of the relevant system~\cite{Adams2008,Cook2009,Cook2009a,Brackley2010a,Greulich2012,Ciandrini2014,Mohapatra2016,Mohapatra2017}. Only a few studies have considered simultaneous limitation of two resources~\cite{Brackley2012}. In particular, the role of resource limitation has not been explored when two processes with antagonistic actions are concurrently affected by the limited availability of protein.

In this Letter, we study the impact of limitations in the supply of both tubulin and the depolymerizing molecular motor Kip3 on the regulation of MT length. We build on a recently validated quantitative model of MT dynamics~\cite{Melbinger2012}, and extend it to include the constraint of resource limitations. We find that Kip3 can tightly control MT length, irrespective of the specific parameter choice. Over a broad range of tubulin and kinesin concentrations, length regulation is bistable, i.e., the MT can assume one of two stationary states. We corroborate these findings by performing \emph{in vitro} experiments, which show that the MT length distribution is indeed bimodal for certain concentrations of the components of interest, in accordance with the theoretical expectations. 

\begin{figure}
	\centering
	\includegraphics[width=.85\columnwidth]{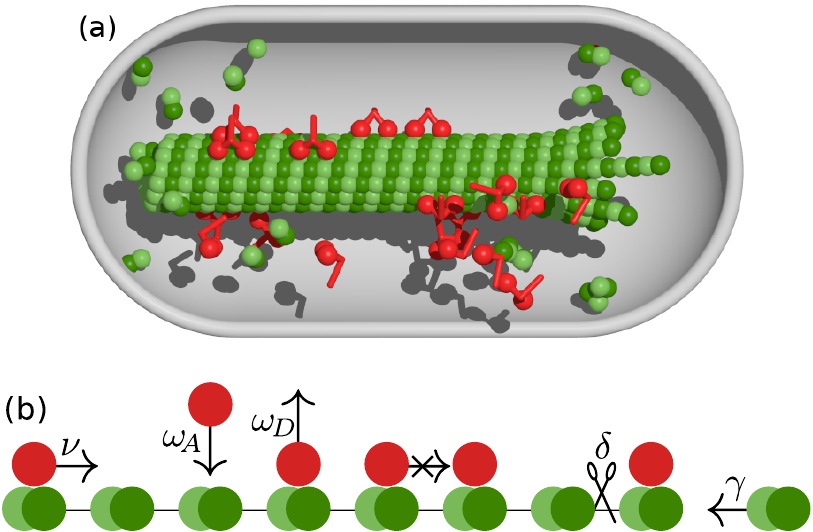}
	\subfloat{\label{sf_model_schematic}}\subfloat{\label{sf_model_lk}}
	\caption{
		Sketch of the model. \protect\subref{sf_model_schematic} A MT in a closed volume interacts with molecular motors. \protect\subref{sf_model_lk} Motors attach to the MT lattice at rate $\omon$, and detach at rate $\omoff$. Motors proceed stepwise toward the plus end at rate $\nu$, provided the next site is unoccupied. At the tip, motor-induced lattice depolymerization (rate $\delta$) competes with spontaneous polymerization (rate $\gamma$). $\omon$ and $\gamma$ and  depend on the concentrations of the proteins available in the closed volume, Eqs.~(\ref{eq_resourcedependence}).
		\label{fig_model_lk}
	}		
\end{figure}
To investigate the impact of limited resources on MT dynamics, we employ a driven diffusive lattice gas model~\cite{Krapivsky2010,Chou2011,Ebbinghaus2012} for spontaneous MT polymerization and kinesin-catalyzed MT depolymerization~\cite{Melbinger2012,Reese2014}, as illustrated in Fig.~\ref{fig_model_lk}. 
As kinesin-8 motors predominantly move along single protofilaments~\cite{Bormuth2012,Bugiel2015}, it suffices to consider a one-dimensional lattice of dynamic length $L(t)$.
The state of each site, $i$, is described by its occupation number, $n_i$, where $n_i=0$ and $n_i=1$ signify an empty and occupied site, respectively. On the MT lattice the dynamics follow the totally asymmetric simple exclusion process with Langmuir kinetics (TASEP/LK)~\cite{Lipowsky2001,Klumpp2003a,Parmeggiani2003,Parmeggiani2004}: Motors can attach to any empty site on the MT lattice at rate $\omon$, and detach at rate $\omoff$. Since binding of motors to the MT depletes the volume concentration of motors $c_m$, the attachment rate $\omon$ decreases as 
\begin{subequations}		
\begin{align}
	\omon = \omon^0 (c_m - m/V) ~.
	\label{eq_omon}
\end{align} 
Here $m$ is the number of motors attached to a protofilament, and $V$ is the effective volume available to the motors, see Sec.~S.III in the Supp.~Mat.~\cite{Supplement}. We are specifically interested in the molecular motor Kip3~\cite{Varga2006,Gupta2006}, which is the kinesin-8 homolog in \emph{S. cerevisiae}. Based on published \emph{in vitro} single-molecule experiments, we estimate its detachment rate to be $\omoff = 4.9 \cdot 10^{-3}~\s^{-1}$ and the attachment rate to any vacant site as $\omon^0=6.7 \cdot 10^{-4}~\nM^{-1} ~\s^{-1}$~\cite{Varga2009}; see Sec.~S.III. On a protofilament, motors move toward the plus end at rate $\nu=6.35~\s^{-1}$ provided that the next site is empty~\cite{Varga2009}. At the plus end, Kip3 catalyzes MT shrinkage~\cite{Walczak2013}. This is described as a stochastic process where a motor arriving at the last site removes it at rate $\delta=2.3~\s^{-1}$~\cite{Varga2009}. At the same time, MTs polymerize spontaneously through attachment of single tubulin heterodimers to their plus ends. As tubulin resources are limited, this decreases the volume concentration of tubulin $c_T$ and the polymerization rate, 
\begin{align} 
	\gamma= \gamma^0 [ c_T-L/(aV) ]~,
	\label{eq_gamma}
\end{align}
\label{eq_resourcedependence}
\end{subequations} decreases with increasing MT length; here, $a=8.4~\nm$ is the size of a tubulin dimer~\cite{Hyman1995}, the (net) polymerization rate per protofilament is $\gamma_0=0.38~\muM^{-1} \s^{-1}$~\cite{Hyman1992,Brouhard2008a}, 
and the effective volume is $V \approx 1.66~\mum^3$ (Sec.~S.III).
\begin{figure}
	\centering
	\includegraphics[height=.182\textwidth]{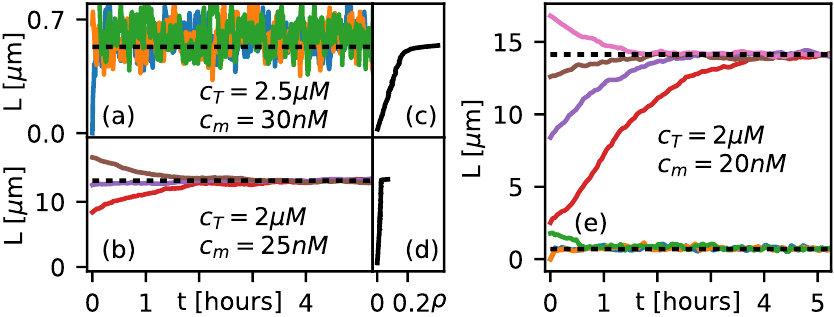}
	\subfloat{\label{sf_trajectories_a}}\subfloat{\label{sf_trajectories_b}}\subfloat{\label{sf_densprof_a}}\subfloat{\label{sf_densprof_b}}\subfloat{\label{sf_trajectories_bistability}}
	\caption{
		Basic phenomenology. Differently colored traces depict different simulation runs under the indicated starting conditions. \protect\subref{sf_trajectories_a}--\protect\subref{sf_trajectories_b} MTs evolve toward a stationary length, which depends on the concentrations of Kip3 and tubulin. \protect\subref{sf_densprof_a}--\protect\subref{sf_densprof_b} The corresponding motor density $\rho$ on the MT is shown; see Fig.~S2e: $\rho$ increases with distance from the minus end, and peaks at the plus end. \protect\subref{sf_trajectories_bistability} The MT length is bistable for a range of concentrations. Dotted lines: Results of the full mean-field theory.
		\label{fig_trajectories}
	}		
\end{figure}

We performed extensive stochastic simulations~\cite{Gillespie1977a}, and explored how the MT dynamics depend on the volume concentration of the motor Kip3, $c_m$, and tubulin, $c_T$. Figure~\ref{sf_trajectories_a} and \ref{sf_trajectories_b} show the dynamics of MT length for two representative concentrations of both components, and for various initial MT lengths. In all cases, the MT length reaches a stationary state, albeit at different values. Moreover, the corresponding motor density $\rho$ also differs (Figs.~\ref{sf_densprof_a}--\ref{sf_densprof_b}): While for a short stationary length, the overall motor density is relatively high, it remains low when the MT length is long. For both cases, the motor density peaks at the plus end (forming a \enquote{spike}). 

\begin{figure*}
	\centering
	\includegraphics[height=.184\textwidth]{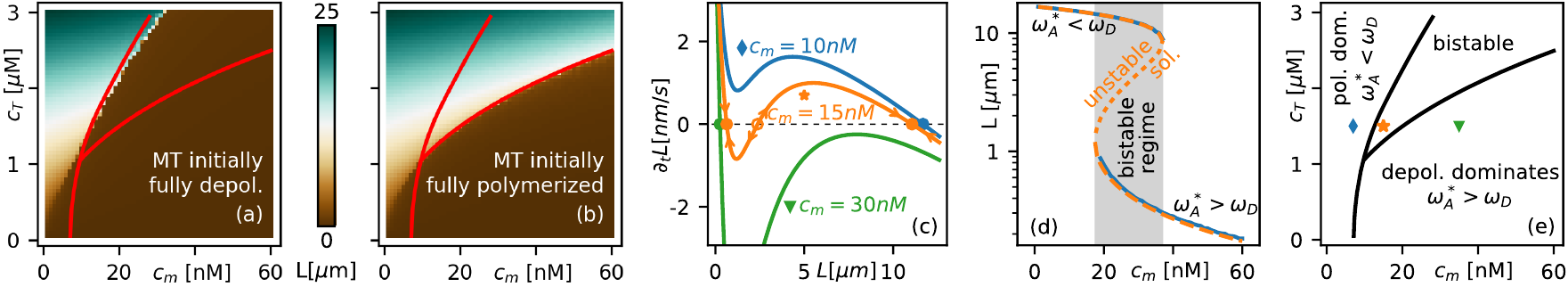}
	\subfloat{\label{sf_parameter_scan_short}}\subfloat{\label{sf_parameter_scan_long}}\subfloat{\label{sf_stability}}\subfloat{\label{sf_comparison_theory_sims}}\subfloat{\label{sf_pd_theory}}
	\caption{
		Theoretical results. \protect\subref{sf_parameter_scan_short}--\protect\subref{sf_parameter_scan_long} \emph{In silico} scans of the stationary length of MTs, shown in color, as a function of $c_m$ and $c_T$. Simulations start from a fully depolymerized lattice (short) in panel \protect\subref{sf_parameter_scan_short}, in \protect\subref{sf_parameter_scan_long} the MT is initially fully polymerized (i.e., long). In the region bounded by the red lines (obtained from the full MF theory, Sec.~S.II), the stationary length differs for these two cases: Here, MT dynamics is bistable. 
		\protect\subref{sf_stability} Rate of change of the MT length, $\partial_t L$, as a function of $L$ at $c_T=1.5~\muM$ for three different motor concentrations, as obtained from the approximate MF theory. For low and high motor concentrations, MT length is monostable, while for intermediate concentrations, two stable stationary states are separated by an unstable state (bistability). \protect\subref{sf_comparison_theory_sims} Comparison of the steady-state length obtained from simulations (blue) and the full MF theory (orange) at $c_T=2~\muM$. \protect\subref{sf_pd_theory} Stability diagram as obtained from the full MF theory.
		\label{fig_parameter_scan}
		\label{fig_bistability_theory}
	}
\end{figure*}

We observe that the stationary MT length depends on its initial value for a certain range of tubulin and kinesin concentrations. Here, depending on whether the MT starts from a single tubulin dimer or a fully polymerized filament with all tubulin resources depleted, the stationary length is short or long, respectively, i.e. the MT dynamics is bistable (Fig.~\ref{sf_trajectories_bistability}).
This is fundamentally different from MT length regulation with unlimited resources~\cite{Melbinger2012,Reese2014}, where only one stationary state of finite length is observed. A parameter scan of the tubulin and Kip3 concentrations (Figs.~\ref{sf_parameter_scan_short}--\ref{sf_parameter_scan_long}) shows that bistable length regulation occurs over a broad parameter range.

What physical processes determine MT length and lead to bistability? To answer this question one needs to analyze the intricate interplay between the crowding of molecular motors in the lattice bulk~\cite{Lipowsky2001,Parmeggiani2004} and the (de)polymerization kinetics at the MT tip~\cite{Reese2011}, as well as the exchange of resources between filament and solution. The rate of change of protofilament length is determined by the antagonism between spontaneous polymerization and Kip3-driven depolymerization kinetics, 
\begin{subequations}
\begin{align}
	\partial_t L(t) = \bigl[ \gamma(t)-\rho_+(t) \delta \bigr] a~, 
	\label{eq_dtL_lk}
\end{align}
where $\rho_+$ is the probability that the terminal site, i.e. the site directly at the MT tip, is occupied by a motor. The number of motors on the protofilament changes when a motor attaches to one of the empty lattice sites, or any of the  motors on it detaches; the number also decreases when a motor falls off the plus end, taking the last tubulin heterodimer with it. Together, this yields 
\begin{align}
	\partial_t m(t)\!=\omon(t)\!\bigl[ L(t)/a-m(t) \bigr]\!- \omoff m(t)\!- \rho_+(t) \delta 
	\label{eq_dtm_lk}.
\end{align}
\label{eq_balance}
\end{subequations}
In Eqs.~(\ref{eq_balance}), the tip density $\rho_+$ drives the loss of tubulin dimers and motors due to depolymerization. This density, in turn, is determined by the flux of motors along the protofilament toward the MT tip. We assume that these bulk dynamics are fast in comparison to MT length changes due to polymerization and depolymerization. Given this time scale separation, the bulk density can be assumed to be stationary (Sec.~S.II A), such that the tip density is determined by a balance between bulk current and depolymerization current. Neglecting correlations in the motor density, $\langle n_i n_j \rangle \approx \langle n_i \rangle \langle n_j \rangle$, and imposing a continuum limit, the mean-field (MF) bulk current is given by $j(x)=\nu \rho(x) [1-\rho(x)]$, where $\rho(x)$ denotes the average motor density at position $x$. On length scales of the order of the size of a tubulin dimer $a$, this current is constant since $\omon, \omoff \ll \nu$, such that the motor flux in the MT bulk equals the flux off the tip: $\nu \rho_{L-a}(1-\rho_{L-a}) \approx \rho_+ \delta$.
Here, the subscript $L-a$ signifies that the density is evaluated very close to the MT plus end, just before the density spike begins (cf. Fig.~\ref{sf_densprof_a}--\ref{sf_densprof_b}); note that in general $\rho_+ \neq \rho_{L-a}$. 

In order to determine the bulk density $\rho_{L-a}$, one needs to consider the combined effects of steric exclusion and motor exchange between filament and cytosol along the complete MT. In the stationary state, changes in motor density caused by transport are balanced by attachment-detachment kinetics, i.e., 
\begin{equation}
	\nu a \bigl( 2 \rho-1 \bigr) \partial_x \rho= -\omon (1-\rho) +\omoff \rho~.
	\label{eq_dxrho}
\end{equation}
This differential equation has solutions in terms of Lambert $W$-functions~\cite{Parmeggiani2003,Parmeggiani2004} which allow one to compute $\rho_{L-a}$ without any further approximations (Sec.~S.II). However, much can already be learned from an approximate solution, where the density is approximated as a Taylor series, $\rho(x) \approx A x+Bx^2$; note that $\rho(0)=0$. Upon inserting this expression into Eq.~(\ref{eq_dxrho}), $A$ and $B$ can be read off by comparing the coefficients in the ensuing power series, and using $\rho_{L-a} \approx \rho(L)$. The motor current off the MT, $\rho_+ \delta$, is now readily computed, and one obtains to second order in $\omega_{\textnormal{A,D}} L/a$:
\begin{equation}
	\rho_+ \delta \approx \omon L/a - (\omon +\omoff ) (L/a)^2 \omon / (2\nu ) ~. \\
	\label{eq_endfluxapprox}
\end{equation}
With Eqs.~(\ref{eq_balance}) and (\ref{eq_endfluxapprox}), we have arrived at a closed set of (nonlinear) equations for the dynamics of the MT length  and the number of motors  bound to a protofilament. It can be viewed as a dynamical system which, as a function of the control parameters $c_m$ and $c_T$, may show bifurcations in the number and nature of its steady states.

The dynamics of nonlinear systems is best visualized by the flow field $(\partial_t m, \partial_t L)$ in phase space. Here, the MT state, described by $L$ and $m$, evolves along the lines drawn in a stream plot (Figs.~S17). This analysis shows that the number of motors bound to the MT equilibrates almost instantaneously, much more rapidly than the MT length changes. Therefore, we can assume that the dynamics reduces to the subspace (nullcline) $\partial_t m=0$. This adiabatic elimination of $m$ yields an effective dynamics of the MT length $L(t)$, as shown in Fig.~\ref{sf_stability}. Keeping the tubulin concentration fixed at a typical value of $1.5 ~\muM$, we find that if the motor concentration is either low ($c_m=10~\nM$) or high ($c_m=30~\nM$)), there is only a single state where the MT length becomes stationary. Hence, regardless of its initial length, a MT will always reach a uniquely defined stable steady length (monostability). By contrast, for intermediate motor concentrations ($c_m=20~\nM$), we observe bistability: Here, three stationary states exist, two stable states for long and short MT lengths, respectively, and one unstable state at intermediate MT length, see Fig.~\ref{sf_stability}. This implies that, depending on its initial length, a MT may either grow long or remain short. The same behavior is observed for the full MF analysis, which includes an exact solution of Eq.~(\ref{eq_dxrho}); see Sec.~S.II.

Figure~\ref{sf_comparison_theory_sims} shows that the results obtained from the full MF theory compare very well with those of the stochastic simulations. In particular, we consistently observe a bistable regime, with two stable solutions separated by an unstable solution (separatrix). The stability diagram shown in Fig.~\ref{sf_pd_theory} summarizes the different regimes of length regulation as a function of protein concentrations. In the regimes dominated by depolymerization or polymerization, the stationary MT length will be short or long, respectively. At intermediate protein concentrations, the MT length may be short or long depending on the initial length (bistable regime). 

\begin{figure}
	\centering
	\subfloat{\includegraphics[width=.99\columnwidth]{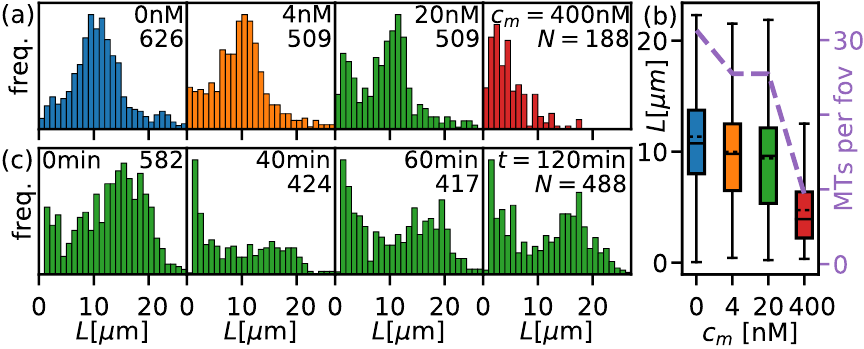}\label{sf_exp_histograms}}\subfloat{\label{sf_exp_box}}\subfloat{\label{sf_varying_inc_time}}
	\caption{ 
		Experimental results. \protect\subref{sf_exp_histograms} Length distribution of MTs grown for $3$ hours and subsequently incubated with various concentrations of Kip3 for one hour. The distribution is unimodal for $0~\nM$, $4~\nM$, and $400~\nM$ Kip3; it is bimodal for $20~\nM$ Kip3, indicating that length regulation is bistable. 
		\protect\subref{sf_exp_box} Box plots for the MT length show that the median MT length decreases as the Kip3 concentration is increased (left axis). The dashed line (right axis) indicates the average number of MTs per field of view (fov).
		\protect\subref{sf_varying_inc_time} Length distributions of MTs initially grown for 3 hours and subsequently incubated with $20~\nM$ Kip3 for various amounts of time. Within an hour, a bistable distribution is established, and its shape is conserved as the incubation time is increased.
		\label{fig_experiment}
	}
\end{figure}
While these results have been obtained for a single MT, they are not limited to this case. 
We find that when many MTs globally access proteins in a well-mixed pool, length regulation is still accurate, see Sec.~S.V.
Moreover, the total length of MTs is bistable in a concentration regime similar to the single-MT case, Fig.~S8.
Here, all MTs jointly become short or long, and their average initial length determines which of these states is reached.

Because diffusion in a real system is fast only on short length scales, and large systems are not well-mixed, we decided to test the actual behavior directly by performing a set of \emph{in vitro} experiments. We grew GMP-CPP stabilized MTs from a MT polymerization solution based on $2~\muM$ tubulin at 27 $^\circ$C (see Sec.~S.VI for details). The resulting MTs had a length distribution similar to a Schulz distribution~\cite{Jeune-Smith2010}, and their median length could be influenced by varying the incubation time ($1.5$ or $3$ hours). Subsequent to initial MT polymerization, different Kip3 concentrations were added to samples of the same polymerization solution and, as a control, no Kip3 was added to the final aliquot; all parts were incubated for another hour, so that MT polymerization from the remaining free tubulin and Kip3-induced depolymerization could occur simultaneously. The resultant MT length distributions in the samples were imaged as described in Sec.~S.VI.

In the first experiment, MTs grown for $3$ hours were incubated with 0, 4, 20, and 400$~\nM$ Kip3 for another hour. In the absence of Kip3 (0 nM), the length distribution of MTs peaked around $11~\mum$ (Fig.~\ref{sf_exp_histograms} and Fig.~S20). The presence of Kip3 reduced the median MT length (Fig.~\ref{sf_exp_box}, box plots/left axis), and also decreased the number of MTs per field of view (Fig.~\ref{sf_exp_box}, dashed line/right axis). The latter indicates that a number of MTs were completely depolymerized or shrank below the detection limit of our setup. Notably, at low and high Kip3 concentrations of $4~\nM$ and $400~\nM$, the length distributions were unimodal with peaks around $11~\mum$ and $2.5~\mum$ (Fig.~\ref{sf_exp_histograms}), respectively. This corresponds to the monostable regimes at low and high motor concentrations derived above, where polymerization and depolymerization, respectively, dominate and the final length is independent of the initial length. 

In contrast, at a Kip3 concentration of $20~\nM$, the MT length distribution was qualitatively different: Here, two distinct populations of MTs (peaks around $2.5~\mum$ and $11~\mum$) were observed, resulting in a bimodal length distribution (Fig.~\ref{sf_exp_histograms}).
We could exclude that the short MTs observed in this experiment were additionally nucleated after the addition of Kip3, Fig.~S19.
Furthermore, the two peaks are not transient. In contrast, a bimodal distribution is fully established within an hour, and the qualitative distribution remained intact until the end of our experiment, Fig.~\ref{sf_varying_inc_time}.
The bimodal length distribution must therefore result from bistable length regulation:
According to Fig.~S7, MTs in a well-mixed many-filament system will jointly become short or long in the bistable regime, and their local average initial length distinguishes between these cases. 
However, diffusion of protein is fast only on short length scales, and slow in large systems, such as our experimental setup.
In addition, the association of motors with MTs, on which they may remain for minutes or longer, significantly slows down diffusion in crowded environments, e.g., inside cells.
The resulting separation of length scales of the small well-mixed range and the large system size may hence allow different regions of a system to develop independently.
Given a broad initial distribution of MT lengths, the local average length of MTs in some regions is therefore in the domain of attraction of the steady state with long length, while in other regions MTs are attracted towards the short length.
Hence, MTs in distant spatial regions evolve towards the different fixed points and domains with long and short filaments are formed, which coexist at stationarity.
This interpretation is supported by the length distribution of MTs resulting from a solution of Kip3 and tubulin which is incubated for 1 hour in a shaker at the same conditions otherwise, Fig.~S18.
Because constant mixing leads to a global well-mixed reservoir, the resulting length distribution is unimodal, confirming our expectations.

We then sought to obtain further information about the domains of attraction of the respective stationary states and the corresponding separatrix marking the boundary between these domains (Fig.~S17b). 
If the MT length distribution at which the length regulation process starts is short, MTs in all regions will be in the domain of attraction of the short stationary length.
To test this prediction, we stopped MT growth after $1.5$ hours and subsequently added the same amounts of Kip3 to the polymerization solution as before. The median MT length in the absence of Kip3 was significantly shorter (Fig.~S15b) than the corresponding value for MTs grown for $3$ hours. We observed that the length distribution remained unimodal when Kip3 was added, irrespective of its concentration, Fig.~S15a. This indicates that, after 1.5 hours of initial MT polymerization, filament lengths still lie below the separatrix in Fig.~S15b. Taken together, our experimental findings qualitatively confirm our theoretical predictions, including the existence of a regime where MT length regulation by Kip3 gives rise to two populations of filaments with clearly distinct lengths.

Taking a broader perspective, we believe that -- similar to the case considered here -- effects of resource limitation are of relevance to other aspects of mitotic spindle formation and disassembly, and other processes in which protein availability in the cytosol constrains dynamic interactions. 

\begin{acknowledgments}
	E.F. and S.D. acknowledge support by the German Excellence Initiative via the programs \enquote{NanoSystems Initiative Munich} (NIM) and \enquote{Center for Advancing Electronics Dresden} (cfaed), respectively.
	L.R. acknowledges support by the Netherlands Organisation for Scientific Research (NWO), via the researc program \enquote{Membrane-cytoplasm protein shuttling}, Project No. 13PMP03.
\end{acknowledgments}

\cleardoublepage
\pagenumbering{gobble}
\begin{tikzpicture}[remember picture,overlay,shift={(current page.center)}]
\node[inner sep=0pt] at (0,0) {\includegraphics[width=\paperwidth,page=1]{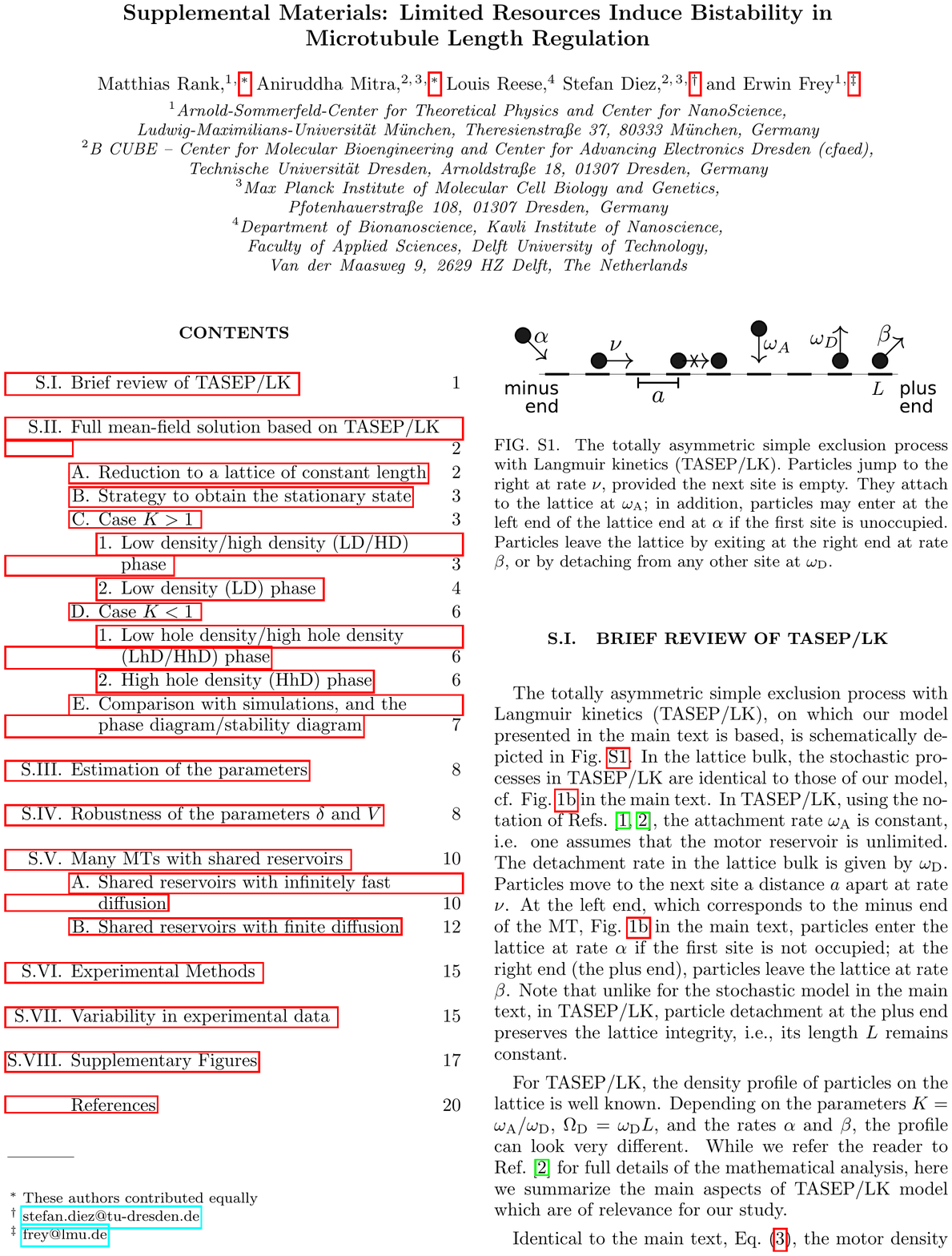}};
\end{tikzpicture}

\cleardoublepage
\begin{tikzpicture}[remember picture,overlay,shift={(current page.center)}]
\node[inner sep=0pt] at (0,0) {\includegraphics[width=\paperwidth,page=2]{supplement_final.pdf}};
\end{tikzpicture}

\cleardoublepage
\begin{tikzpicture}[remember picture,overlay,shift={(current page.center)}]
\node[inner sep=0pt] at (0,0) {\includegraphics[width=\paperwidth,page=3]{supplement_final.pdf}};
\end{tikzpicture}

\cleardoublepage
\begin{tikzpicture}[remember picture,overlay,shift={(current page.center)}]
\node[inner sep=0pt] at (0,0) {\includegraphics[width=\paperwidth,page=4]{supplement_final.pdf}};
\end{tikzpicture}

\cleardoublepage
\begin{tikzpicture}[remember picture,overlay,shift={(current page.center)}]
\node[inner sep=0pt] at (0,0) {\includegraphics[width=\paperwidth,page=5]{supplement_final.pdf}};
\end{tikzpicture}

\cleardoublepage
\begin{tikzpicture}[remember picture,overlay,shift={(current page.center)}]
\node[inner sep=0pt] at (0,0) {\includegraphics[width=\paperwidth,page=6]{supplement_final.pdf}};
\end{tikzpicture}

\cleardoublepage
\begin{tikzpicture}[remember picture,overlay,shift={(current page.center)}]
\node[inner sep=0pt] at (0,0) {\includegraphics[width=\paperwidth,page=7]{supplement_final.pdf}};
\end{tikzpicture}

\cleardoublepage
\begin{tikzpicture}[remember picture,overlay,shift={(current page.center)}]
\node[inner sep=0pt] at (0,0) {\includegraphics[width=\paperwidth,page=8]{supplement_final.pdf}};
\end{tikzpicture}

\cleardoublepage
\begin{tikzpicture}[remember picture,overlay,shift={(current page.center)}]
\node[inner sep=0pt] at (0,0) {\includegraphics[width=\paperwidth,page=9]{supplement_final.pdf}};
\end{tikzpicture}

\cleardoublepage
\begin{tikzpicture}[remember picture,overlay,shift={(current page.center)}]
\node[inner sep=0pt] at (0,0) {\includegraphics[width=\paperwidth,page=10]{supplement_final.pdf}};
\end{tikzpicture}

\cleardoublepage
\begin{tikzpicture}[remember picture,overlay,shift={(current page.center)}]
\node[inner sep=0pt] at (0,0) {\includegraphics[width=\paperwidth,page=11]{supplement_final.pdf}};
\end{tikzpicture}

\cleardoublepage
\begin{tikzpicture}[remember picture,overlay,shift={(current page.center)}]
\node[inner sep=0pt] at (0,0) {\includegraphics[width=\paperwidth,page=12]{supplement_final.pdf}};
\end{tikzpicture}

\cleardoublepage
\begin{tikzpicture}[remember picture,overlay,shift={(current page.center)}]
\node[inner sep=0pt] at (0,0) {\includegraphics[width=\paperwidth,page=13]{supplement_final.pdf}};
\end{tikzpicture}

\cleardoublepage
\begin{tikzpicture}[remember picture,overlay,shift={(current page.center)}]
\node[inner sep=0pt] at (0,0) {\includegraphics[width=\paperwidth,page=14]{supplement_final.pdf}};
\end{tikzpicture}

\cleardoublepage
\begin{tikzpicture}[remember picture,overlay,shift={(current page.center)}]
\node[inner sep=0pt] at (0,0) {\includegraphics[width=\paperwidth,page=15]{supplement_final.pdf}};
\end{tikzpicture}

\cleardoublepage
\begin{tikzpicture}[remember picture,overlay,shift={(current page.center)}]
\node[inner sep=0pt] at (0,0) {\includegraphics[width=\paperwidth,page=16]{supplement_final.pdf}};
\end{tikzpicture}

\cleardoublepage
\begin{tikzpicture}[remember picture,overlay,shift={(current page.center)}]
\node[inner sep=0pt] at (0,0) {\includegraphics[width=\paperwidth,page=17]{supplement_final.pdf}};
\end{tikzpicture}

\cleardoublepage
\begin{tikzpicture}[remember picture,overlay,shift={(current page.center)}]
\node[inner sep=0pt] at (0,0) {\includegraphics[width=\paperwidth,page=18]{supplement_final.pdf}};
\end{tikzpicture}

\cleardoublepage
\begin{tikzpicture}[remember picture,overlay,shift={(current page.center)}]
\node[inner sep=0pt] at (0,0) {\includegraphics[width=\paperwidth,page=19]{supplement_final.pdf}};
\end{tikzpicture}

\cleardoublepage
\begin{tikzpicture}[remember picture,overlay,shift={(current page.center)}]
\node[inner sep=0pt] at (0,0) {\includegraphics[width=\paperwidth,page=20]{supplement_final.pdf}};
\end{tikzpicture}

\end{document}